\documentclass[12pt]{iopart}
\pdfoutput=1
\usepackage{graphicx}
\usepackage{placeins}
\usepackage{float}
\usepackage{subfigure}
\begin{document}
\title[arXiv.org]{An improved electron beam dynamics design for laboratory plasma-astrophysical studies: a technical note\footnote{The note is supplementary to the framework of laboratory plasma-astrophysical studies previously carried out at Deutsches Elektronen-Synchrotron in Refs.\ [3-5].}}
\author{Ye Chen and Chun-Sung Jao}
\ead{ye.lining.chen@gmail.com}
\begin{abstract}A technical note is given regarding our previous laboratory plasma-astrophysical studies [C.-S. Jao et al., High Energy Density Physics 32, 31-43 (2019) and Y. Chen et al., Nucl. Instrum. Methods Phys. Res., Sect. A 903, 119 (2018)]. In this note, an upgraded accelerator beamline design is proposed based on a feasible experimental setup in a realistic laboratory environment. The improved design aims to provide milliampere (mA) mega-electron-volt (MeV) quasi-continuous (cw) electron beams for plasma-astrophysical applications. Such a design utilizes a so-called mixed-guiding-field magnetic system right after the cut disk structure (CDS) booster cavity to provide a periodic longitudinal focusing field. The transportation of the produced cw beam with large energy spread to the plasma cell location is improved. The magnetic field serves as well as a seeding field in the plasma environment for the growth of electromagnetic instabilities. In conjunction with the appliance of a circular collimator at the exit of the CDS, the new design allows production of quasi-cw beams with a three orders higher number density at the entrance of the plasma cell compared to the previous design for a seeding magnetic field of about 50 mT while the locally enhanced electric field at the cathode is up to 8 GV/m. The associated beam dynamics simulation results are presented. As proof of principle studies, the produced electron beams are applied in nonlinear plasma-astrophysical simulations for exploring the growth of the instabilities. The extracted parameters and/or distributions from the generated electron beams in the laboratory environment are used in these particle-in-cell simulations. The obtained results are presented and discussed.
\end{abstract}
\maketitle
\section{Introduction}
\label{sec:introduction}
Laboratory plasma-astrophysics has became one of the most significant branches engrossing in astrophysics over the last decades. As an alternative method apart from the classical observation and numerical simulation \cite{Overview1, Overview2}, it provides a novel methodology for exploring the mechanisms of  astrophysical phenomena in a realistic laboratory environment based on advanced technologies and modern instrumentation in high energy physics. A research program was launched at Deutsches Elektronen-Synchrotron for establishing a so-called plasma-astrophysical laboratory for experimental studies of beam-plasma instabilities \cite{NIMA:Ye, Prestudy, Prestudy2}.

With feasible experimental setups in such a laboratory environment, particle beams need to be produced in a controlled manner to fulfill a set of compromised requirements between several crucial electron beam parameters for developing the instabilities. This does include not only the duration (preferably continuous), the average current ($\geq$mA), the average energy (several MeV at least), the number density, but also other instability-mechanism-oriented quality properties of the required particle beam. In principle, a long beam (continuous) of high average current (mA) is required as the energy source to excite the plasma wave. A higher beam energy is expected to induce stronger perturbation of the instability due to the fact that plasma wave excitation is led by the drift energy of streaming particles. Higher beam energy ($\geq$MeV) results also in a higher growth rate of the instability. From this point of view the appliance of RF injectors for our purpose can be much more beneficial due to higher beam energy and principally better beam qualities compared to DC guns \cite{Book:FS, Mikhail:PRAB2012}. A matched number density of the particle beam with the plasma density is of another crucial importance to grow the instability. This requires specially designed beam transport system for properly focusing the beam with large energy spread at the plasma cell location while maintaining other beam quality parameters as required. 

A preliminary design was reported in \cite{NIMA:Ye}. It demonstrated electron bunch extraction based on field emission from a designed metallic needle cathode and quasi-cw beam formation using velocity bunching over subsequent radio-frequency (RF) cycles of a cut disk structure (CDS) downstream of an L-band RF electron gun. In this note, an improved beam dynamics design is proposed and will be further discussed based on 3D particle tracking simulations (Sec.\ \ref{BeamDynamics}). Based on the improved design, the quasi-cw beams with fulfilled requirements produced in a realistic laboratory environment are applied in nonlinear plasma-astrophysical simulations for studying the beam-plasma instabilities. This involves the electrostatic streaming instability \cite{Estatic1, Estatic2, Estatic3},  the filamentation instability \cite{Fila1, Fila2} and a non-resonant streaming instability (i.e.\ Bell's instability\cite{Bells}). Among those instabilities, the electrostatic instability is commonly known as the unstable mode with the shortest wavelength and the shortest development time \cite{Beamins}. As a competition mode, the prior existence of the electrostatic instability is critical to the laboratory observation of other aforestated instabilities, in terms of its parametric dependencies on the realistic particle beams produced in the laboratory and thereby plausible suppression schemes for it. The obtained results consisting of case studies for the so-called cold beam, warm beam and quasi-cw beam will be shown in this note (Sec.\ \ref{BeamPlasma}). A summary and an outlook will be given in Sec.\ \ref{Summary}.

\section{Design of electron beam dynamics}\label{BeamDynamics}

\begin{figure}[!htb]
\centering
\includegraphics[width=140mm,height=30mm]{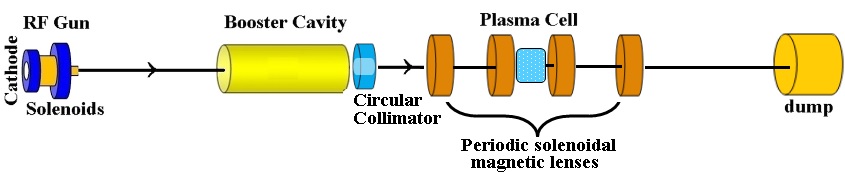}
\caption{A proposed laboratory experimental setup for the enhancement of electron number density in the plasma cell.}
\label{Beamline}
\end{figure}

Figure \ref{Beamline} shows a proposed experimental setup. It consists of a specially designed needle cathode, an L-band RF injector \cite{Book:FS}, a pair of focusing magnets,  a cut disk structure (CDS), a circular collimator at the exit of the CDS, a periodic solenoidal focusing system, a plasma cell and a beam dump. The field-emitted (FE) electron bunch is generated from the needle cathode sitting on the backplane of the 1.3 GHz copper resonator through highly enhanced electric field gradient up to 8 GV/m. The electron bunches produced over subsequent RF cycles are accelerated by the RF gun and velocity debunched through the 1.3 GHz CDS forming a quasi-cw beam by jointing existing gaps in the time domain profile of the produced FE beam between neighbouring RF periods. A set of required beam parameters are produced in \cite{NIMA:Ye}, however, the scheme of beam focusing is not efficient for transporting the produced particle beam to the plasma cell location due to large beam energy spread. Consequently, the resulting electron number density when transported to the entrance of the plasma cell location are limited to approx.\ $10^7-10^8$/cm$^3$ for an enhanced electric field gradient of 5.7 GV/m at the cathode.

As shown in Fig.\ \ref{Beamline}, a periodic magnetic focusing system is applied. Such a system provides a longitudinal mixed magnetic field formed by overlapping fields of an array of solenoidal lenses (e.g. four solenoids exemplarily shown in Fig.\ \ref{Beamline}). A mathematical description of the periodical field on the longitudinal axis can be represented using a Fourier series \cite{EPAC:Militsyn}
\begin{equation}
Bz(z) = B_0[1+\Sigma_na_n cos(k_0z)],
\end{equation}where $Bz$ represents the mixed guiding magnetic field in the longitudinal (z) direction and $B_0$ stands for the magnetic field strength of the solenoid. The term $k_0$=2$\pi$/$T$ with $T$ denoting the period of the solenoidal array. The symbol $a_n$ is defined as the Fourier coefficient with $n=1...\infty$. The mixed guiding field $Bz(z)$ is numerically reconstructed by using four solenoidal lenses periodically arranged in the beam propagation direction and used in the following beam dynamics simulations.

\begin{figure}[!htb]
\centering
\includegraphics[width=95mm,height=70mm]{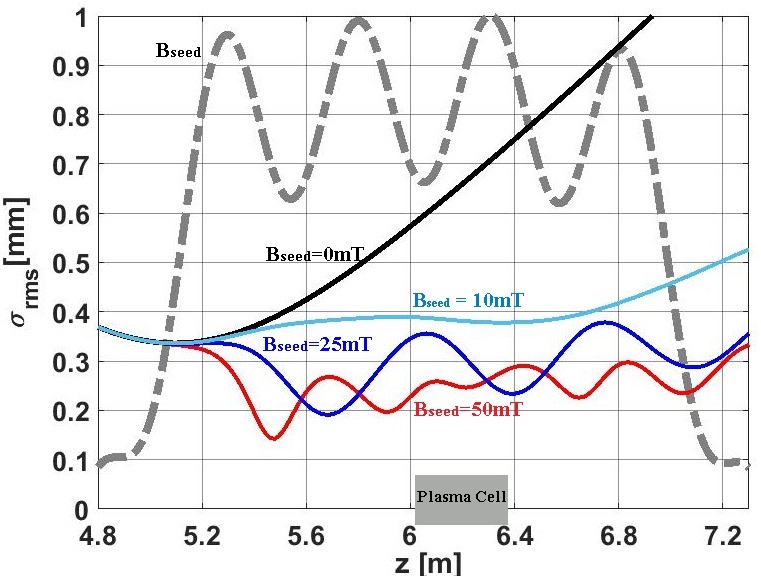}
\caption{Significant improvement of beam focusing within the plasma cell using a periodic magnetic focusing system. $\sigma_{rms}$: rms beam size.}
\label{Bseed}
\end{figure}

Figure \ref{Bseed} shows the rms size of the particle beam in the longitudinal direction with the gray diagram indicating the location of the plasma cell. The evolution of the beam size is compared among multiple cases with the seeding magnetic field $B_{seed}$ (maximum field amplitude of $B_z$) as a variable. The gray dashed curve shows the pattern of the longitudinal magnetic field used for the simulations. The black curve shows in the seeding-field-free case a significant beam size growth within the plasma cell while the ones in cyan, blue and red show a suppressed  behavior of the beam divergence and a much reduced beam size within the plasma cell for an applied seeding field of 10, 25 and 50 mT, respectively. A significant role plays the applied seeding magnetic field in maintaining the beam size and therefore in increasing the electron number density.

\begin{figure}[!htb]
\centering
\includegraphics[width=90mm,height=70mm]{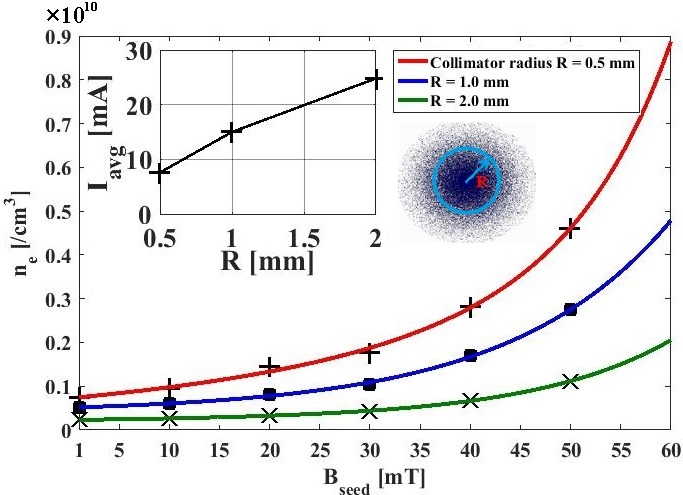}
\caption{Simulated electron number density within the plasma cell as a function of the seeding magnetic field and the collimator radius.}
\label{ND}
\end{figure}

In Fig.\ \ref{ND}, the right inset shows the transverse beam profile right after the CDS where the majority of electrons is concentrated in the central part of the bunching area while for a large beam-halo area much less electrons are populated. Therefore, properly collimating the beam may increase the electron number density while maintaining a sufficiently high beam current. For such a purpose, a circular collimator is applied at the exit of the CDS. Figure \ref{ND} shows the electron number density as a function of the seeding magnetic field (i.e.\ 1-60 mT) with the radius of the collimator as a variable (i.e.\ 0.5, 1.0 and 2.0 mm). The left inset of Fig.\ \ref{ND} shows the average beam current plotted versus the collimator radius. As shown, with a collimator radius of 0.5 mm and a seeding magnetic field of 50 mT, the average beam current can be maintained at about 10 mA for a reached number density on the order of $10^{10}$/cm$^3$.

\begin{figure}[!htb]
\centering
\includegraphics[width=95mm,height=65mm]{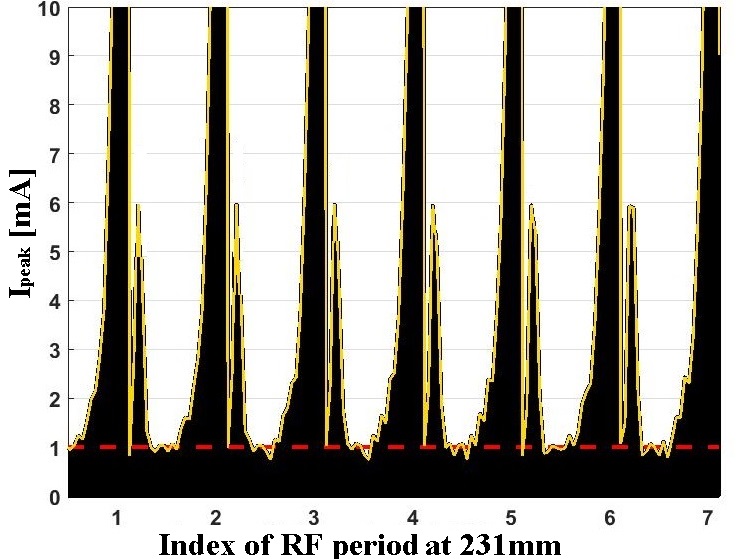}
\caption{Temporal profile of a produced quasi-cw electron beam with a base current of about 1 mA over RF periods at $\lambda=231$ mm.}
\label{CWBeam}
\end{figure}

Furthermore, Fig.\ \ref{CWBeam} shows an exemplary temporal profile of the produced quasi-cw beam based on the improved beam dynamics design. The base current is increased to 1 mA and the electron number density within the plasma cell is on the order of $10^{10}$/cm$^3$. In the following, such electron beams produced in the realistic laboratory environment are used for the nonlinear plasma-astrophysical simulations.

\section{Plasma-astrophysical studies using realistic laboratory electron beam distributions\label{BeamPlasma}}

To understand the physical conditions for beam-plasma instabilities to occur, the extracted parameters and/or distributions of the realistic laboratory electron beams (Sec.\ \ref{BeamDynamics}) are used for plasma-astrophysical simulations. A fully relativistic particle-in-cell code (OSIRIS)  \cite{OSIRIS1, OSIRIS2, OSIRIS3} is employed. Regarding a basic setup for simulations \cite{Prestudy}, a uniform stationary plasma is filled in a two-dimensional open-boundary system ($43.14$~cm~$\times~5.9$~cm) and an electron beam with an initial width of 0.17 cm is applied in the simulation. The number density of stationary plasma is set as $n_0=10^{13}$~cm$^{-3}$. The width of the electron beam is roughly equal to the skin length of the background plasma ($c/\omega_{pe}$ with $\omega_{pe}$ standing for the electron plasma frequency). In exemplary simulations, the injection rate is set as 1.78 GHz instead of the RF frequency of 1.3 GHz based on the assumption that the growth rate of the involved instability can only be slightly varied by changing the injection rate. The settings indicate that the electron bunches are injected into the simulation system every 100 $\omega_{pe}^{-1}$.

\begin{figure}[!htb]
\centering
\includegraphics[width=85mm,height=40mm]{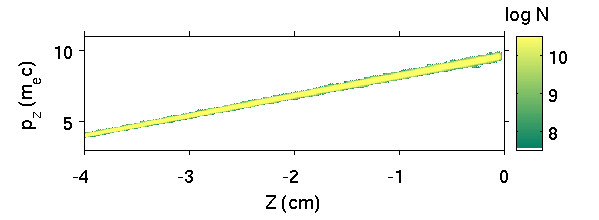}
\caption{The longitudinal momentum distribution of a quasi-cw electron beam imported in PIC simulations.}
\label{BeamProfile}
\end{figure}

Referred to the results from 3D particle tracking simulation, as indicated in Fig.\ \ref{BeamProfile}, a typical momentum distribution of the produced quasi-cw electron beam is shown for a single RF cycle. Due to the time-of-flight effect, a beam momentum shift with a positive slope ($p_z$=10 to 4 $m_ec$ with $m_e$ and $c$ denoting the electron mass and the speed of light, respectively) is clearly observed in the longitudinal direction at the entrance of the plasma cell. A fundamental question arises when a realistic particle distribution is applied in our simulations, in terms of the differences in exciting the streaming instability compared to the conventional assumption of using a cold beam and/or a warm beam. These assumptions essentially refer to the beam temperature which strongly influences the growth rate of the electrostatic streaming instability with the shortest development time among the instabilities of our interest (Sec.\ \ref{sec:introduction}). The linear growth rate of the electrostatic instability, in particular, decreases with increasing the beam temperature\cite{Estatic1, Estatic2, Estatic3, Prestudy, Prestudy2}. Therefore, the impacts of different beam distributions including the produced quasi-cw beam onto the electrostatic instability are worth of first investigations. Nonlinear PIC simulations are thus carried out for three cases, namely, the cold beam case, the warm beam case and the quasi-cw beam case. A homogeneous Maxwell-Boltzmann momentum distribution centering at $7 m_ec$ is set with a thermal velocity ($v_{th,e} $) of 0.01$c$ and 0.2$c$ for the cold and warm beam cases, respectively. For the quasi-cw beam case, a realistic beam momentum distribution is used, as shown in Fig.\ \ref{BeamProfile}.

\begin{figure*}[!htb]
\centering
\includegraphics[width=\textwidth]{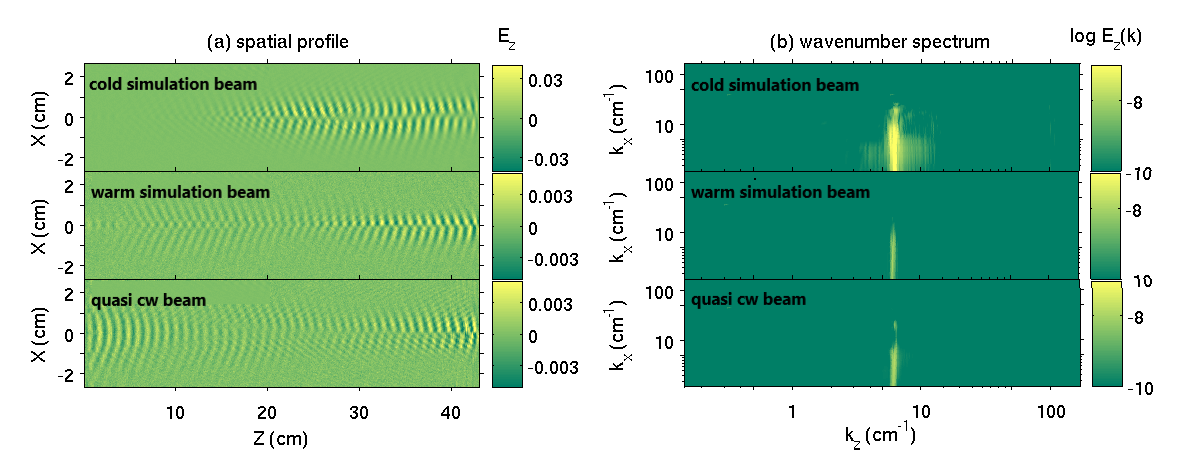}
\caption{Spatial profiles of the longitudinal electric field components (left panels, in units $m_ec\omega_p/e$) and the corresponding Fourier plots (right panels).}
\label{EFProfile}
\end{figure*}

Figure \ref{EFProfile} shows the spatial profile of longitudinal electric field  components (panel a) and the corresponding Fourier plots (panel b) for the three cases at T=5.783 ns (1024 $\omega_{pe}^{-1}$), respectively. As shown, the electric field perturbation is present as the electron beam is propagating through the plasma environment. The major perturbations are along X=0, at which position the electron beam is traveling along a central path with an initial width of 0.17 cm.  The wavelength (wave number) of the major perturbations is about 1.05 cm (5.88 cm$^{-1}$) indicating a reciprocal relation of the wave number
to the skin length of the background plasma. This is well resolved in the PIC simulations. It can therefore be confirmed from the Fourier plots in Fig.\ \ref{EFProfile} (b), that the electrostatic instability occurs in all three cases. Furthermore, the cold beam case shows the strongest electric field perturbation while in the other two cases the instability strengths are on the same order of magnitude and both are much reduced. It should be noted, that the warm beam case turns out to be a good approximation of the quasi-cw beam. Note, in addition, that the appliance of a quasi-cw beam produced in a realistic laboratory environment may be even more beneficial, in terms of the suppression of the electrostatic instability for relieving its competition with the target instability, for studying the beam-plasma instabilities compared to the ideal cold beam case which is typically used in conventional simulations.

\section{Conclusion\label{Summary}}

An improved experimental setup is proposed for studies of beam-plasma instabilities in the laboratory environment. The associated beam dynamics design is carried out to provide quasi-cw electron beams with required parameters. With a circular collimator placed at the exit of the debunching structure, the appliance of a periodic longitudinal magnetic focusing system renders significant improvements in production of mA and MeV quasi-cw electron beams with a three orders higher number density within the plasma cell for a seeding magnetic field of approx. 50 mT, in comparison to the previous design at a conservative working point of the field emitter. 

The produced electron beams are employed in the plasma-astrophysical simulations for investigating the parametric dependencies and thereby the suppression scheme for the competition of the fast-growing electrostatic streaming instability with other instabilities of our interest. Numerical results for multiple cases (i.e.\ cold beam, warm beam and realistic beam) show the well-resolved growing electrostatic instability in the beam-plasma system. The obtained wave numbers well agrees with theoretical predictions. A further comparison of electric field perturbations between the multiple cases demonstrates the strongest instability growth in the cold beam case while comparable growths of the instability for the warm beam and the quasi-cw beam. The results show the warm beam model can be a good approximation  of  the realistic quasi-cw beam, and that the appliance of the quasi-cw beam may be even more beneficial for studying the beam-plasma instabilities in comparison to the ideal cold beam model conventionally used in many theoretical and/or numerical analysis. 


\vspace{10pt}

\begin{thebibliography}{1}
\bibitem{Overview1}
W.E.~Amatucci, URSI Radio Science Bulletin 2006 (319) (2006) 32–66.

\bibitem{Overview2}
Y.~Kuramitsu, Y.~Sakawa, T.~Morita, et al., Plasma Physics and Controlled Fusion 54 (12) (2012) 124049.

\bibitem{NIMA:Ye}
Y.~Chen, G.~Loisch, M.~Gross, et al., Nucl. Instrum. Methods Phys. Res., Sect. A 903, 119 (2018).

\bibitem{Prestudy}
C.-S.~Jao, S.~Vafin, Y.~ Chen, et al., High Energy Density Physics 32, 31-43 (2019).

\bibitem{Prestudy2}
C.-S.~Jao, S.~Vafin, Y.~ Chen, et al., arXiv:1910.13756, 2019.

\bibitem{Book:FS}
F.~Stephan, M.~Krasilnikov, \emph{High Brightness Photo Injectors for Brilliant Light Sources}, in Chapter of the book: Synchrotron Light Sources and Free-Electron Lasers, editors E.~Jaeschke, S.~Khan, J.R.~Schneider, J.B.~Hastings, Springer International Publishing Switzerland, ISBN 978-3-319-04507-8, 2014.

\bibitem{Mikhail:PRAB2012} 
M.~Krasilnikov, F.~Stephan, G.~Asova, et al., Phys. Rev. ST Accel. Beams {\bf 15}, 100701 (2012).

\bibitem{Estatic1}
B.N.~Breizman, Reviews of plasma physics Vol.15, editor: B.B.~Kadomtsev,Consultants Bureau, New York, 1990.

\bibitem{Estatic2}
P.~Chang, A.E.~Broderick, C.~Pfrommer, et al., ApJ. 797, 110 (2014).

\bibitem{Estatic3}
C.-S.~Jao, L.-N.~Hau, Physics of Plasmas 23, 112110 (2016).

\bibitem{Fila1}
R.~Bingham, R.~Short, E.~Williams, et al., Plasma Phys. Control. Fusion 26, 1077 (1984).

\bibitem{Fila2}
P.K.~Shukla, L.~Stenflo, Physics of Fluids B: Plasma Physics 1, 1926 (1989).

\bibitem{Bells}
A.R.~Bell, MNRAS 353, 550-558 (2004).

\bibitem{Beamins}
A.~Bret, L.~Gremillet, M.E.~Dieckmann, Phys. Plasmas. 17, 120501 (2010).

\bibitem{EPAC:Militsyn}
B.L.~Militsyn, C.A.J.~van der Geer, W.H.~Urbanus, et al., \emph{Transport of electron beams with large energy spread in a periodic longitudinal magnetic field}, Proceedings of EPAC 2000, Vienna, Austria, p.1054.

\bibitem{OSIRIS1}
R.A.~Fonseca, L.O.~Silva, F.S.~Tsung, et al., Computational Science -ICCS: International Conference Amsterdam, The Netherlands, April, 2002 Proceedings, Part III, 342-351, 978-3-540-47789-1, 2002.

\bibitem{OSIRIS2}
R.A.~Fonsec, S.F.~Martins, L.O.~Silva, et al., Plasma Physics and Controlled Fusion 50, 12 124034 (2008).

\bibitem{OSIRIS3}
R.A.~Fonseca, J.~Vieira, F.~Fiuza, et al., Plasma Physics and Controlled Fusion 55, 12 124011 (2013).

\end{thebibliography}
\end{document}